\documentclass[12pt]{article}

\usepackage{amssymb}
\usepackage{booktabs}
\usepackage{multirow}
\usepackage{soul}
\usepackage{upgreek}
\usepackage{microtype}
\usepackage{graphicx}
\usepackage{tabularx}
\usepackage{hyperref}

\title{Performance of the RF-detectors of the Astroneu Array}
\author{S. Nonis$^1$, A. Leisos$^1$,
A. Tsirigotis$^1$, I. Gkialas$^2$, K. Papageorgiou$^2$,\\ S. Tzamarias$^3$}
\date{%
$^1$ Physics Laboratory, School of Science and Technology, Hellenic Open University, Patras 26222, Greece \\
$^2$ School of Engineering, Department of Financial and Management Engineering, University of the Aegean, Chios 82100, Greece\\
$^3$ Department of Physics, Aristotle University of Thessaloniki, Thessaloniki 54124, Greece\\
}

\begin{document}
\maketitle

\begin{abstract}

Since 2014, the University Campus of the Hellenic Open University (HOU) hosts the Astroneu array which is dedicated
to the detection of Extensive Air Showers (EAS) induced by high energy Cosmic Rays (CR). The Astroneu array incorporates 9 large particle scintillation detectors and 6 antennas sensitive in the Radio Frequency (RF) range 1-200 MHz. The detectors are adjusted in three autonomous stations operating in an environment with strong electromagnetic background. As shown by previous studies, EAS radio detection in such environments is possible using innovative noise rejection methods, as well as advanced analysis techniques.
In this work we present the analysis of the collected radio data corresponding to an operational period of approximately four years. We present the performance of the Astroneu radio array in reconstructing the EAS axis direction using different RF detector geometrical layouts and a  technique for the 
estimation of the shower core by comparing simulation and experimental data. 
Moreover, we measure  the relative amplitudes of the two mechanisms that give rise to RF emission (Askaryan effect and Geomagnetic emission) and show that they are in good agreement with previous studies as well as with the simulation predictions. 
\end{abstract}

\section{Introduction}\label{intro}
\paragraph*{} Cosmic Rays (CR) is the common term used to describe all radiation formed by relativistic charge particles\footnote{However, by generalizing  the CR definition we can also include high energy gamma rays and neutrinos.} originating from outer space and colliding with the Earth's atmosphere. At low energies (up to $10^5~GeV$), the flux of cosmic rays is mostly of solar origin and large enough to make their direct detection possible by detectors of relatively small surface area deployed outside the atmosphere (in balloons or satellites). At energies beyond the knee ($>10^6~GeV$) the galactic and extra-galactic CR component is characterized by low particle fluxes and the direct detection of CR is not practical. Instead at this energies the study of CR properties is realistic through the analysis of the particle showers induced in the atmosphere. 
 
When a high energy CR reaches the atmosphere, after passing through a certain amount of matter, it eventually collides with an air nucleus and a cascade of secondary particles, the so-called Extensive Air Shower (EAS), is formed. Apart from the particle component of the EAS, during its evolution in the atmosphere a significant amount of electromagnetic radiation is emitted both in optical (fluorescence and Cherenkov light) as well as in the radio frequency (RF) part of the spectrum. Consequently, the most traditional ways to detect EAS are the particle detector arrays (i.e. scintillators or water Cherenkov tanks) deployed at the  ground, as well as optical telescopes that record the atmospheric Cherenkov and fluorescence light emission.
However, in the last twenty two years, a considerable progress has been made concerning the RF detection of EAS (see \cite{1,2} for an extensive review). The EAS radio detection appears quite competitive with  traditional methods for the reconstruction of  the primary particle's main parameters (arrival direction, energy and composition) and  furthermore it is characterized by low-cost detectors as well as  minor dependence on the atmospheric conditions.

There are two fundamental physical processes related to the production of the RF electric field by CR air showers. 
The dominant one is the emission correlated with the geomagnetic field, first introduced by Kahn and Lerche \cite{3}.
The Earth's magnetic field ($\vec{B}_\mathrm{E}$) exerts a Lorentz force on the electrons and positrons of the shower which accelerates them in a direction perpendicular to the EAS axis. As the shower develops, the time variation of the number of electrons and positrons results to RF emission. Moreover,  the generated electric field is polarized in the direction of the Lorentz force ($\vec{n}_\mathrm{sh}\times\vec{B}_\mathrm{E}$), where the propagation direction of the secondary particles can be identified with the direction of the EAS axis, $\vec{n}_\mathrm{sh}$.  Additionally to the geomagnetic mechanism, a subdominant contribution to the emitted electric field originates from the negative charge excess in the EAS front as described by Askaryan \cite{4}. The number of electrons in the EAS front appears increased relative to the number of positrons due to the ionization of the air-molecules caused by secondary shower particles and positron annihilation. In the atmospheric depth where the EAS reaches its maximum number of secondary particles ($X_\mathrm{max}$) an electron excess of 15-25 \% is estimated to appear. The time dependence of this excess  in the shower front produce RF emission with the field vector oriented radially to the EAS axis. The measured electric field on the ground is the superposition of these two contributions and depends strongly on the observer's location with respect to the shower axis.

The Astroneu array \cite{5} is a hybrid EAS detection array deployed at the Hellenic Open University (HOU) campus near the city of Patras in Greece. The array is composed of 3 autonomous stations, each one equipped with both particle and RF detectors for hybrid EAS detection. The particle detectors are large scintillator counters (hereafter noted as Scintillator Detector Module- [SDM]) which were designed and constructed in the Physics Laboratory of HOU \cite{6}, while the RF detectors are dipole antennas produced by the CODALEMA collaboration \cite{7}. Since the area around the stations suffers from intense electromagnetic activity, previous studies \cite{8} have shown that the radio signal from EAS can be successfully detected by imposing an appropriate noise shedding formula. Additionally studies on the RF signal features were implemented by correlating data from both RF and SDM detectors as well as with Monte Carlo (MC) simulations \cite{9}. In these studies, the EAS axis direction was reconstructed using either the RF pulse arrival times or using the pulse power spectrum and innovative signal processing methods \cite{10}.

In this work we present the analysis for the full RF dataset collected for a period of approximately 4 years (2017-2021). The RF system's efficiency in reconstructing the EAS axis direction is evaluated for different geometrical layouts (positions) of the RF detectors. In addition, by determining the polarization of the transmitted electric field, the contribution rate of each mechanism (charge excess and geomagnetic) to the measured RF signal is calculated. In Section \ref{sec1} the station's architecture is shortly reported, while in Section \ref{sec2}  the data sample and the simulation procedure is presented. Section \ref{sec3} describes the analysis related to the estimation of the  EAS axis direction using different RF antenna geometrical layouts and the results are compared with the direction obtained from the SDMs measurements as well as with the simulation predictions. In Section \ref{sec4} a new method for the EAS core reconstruction is presented based on the comparison between RF data and electric field simulations, while in Section \ref{sec5}  the relative strength of the two dominant mechanisms in the measured RF signal amplitude is measured and quantified using the Charge Excess to Geomagnetic Ratio (C$_\mathrm{GRT}$). Furthermore the dependencies of this ratio on the zenith and azimuth angle, as well as on the distance from shower core are investigated. Finally in Section \ref{sec6} conclusions remarks and discussion are quoted.  

\section{The Astroneu Array}\label{sec1}
\begin{figure}[h!]
\includegraphics[width=\textwidth]{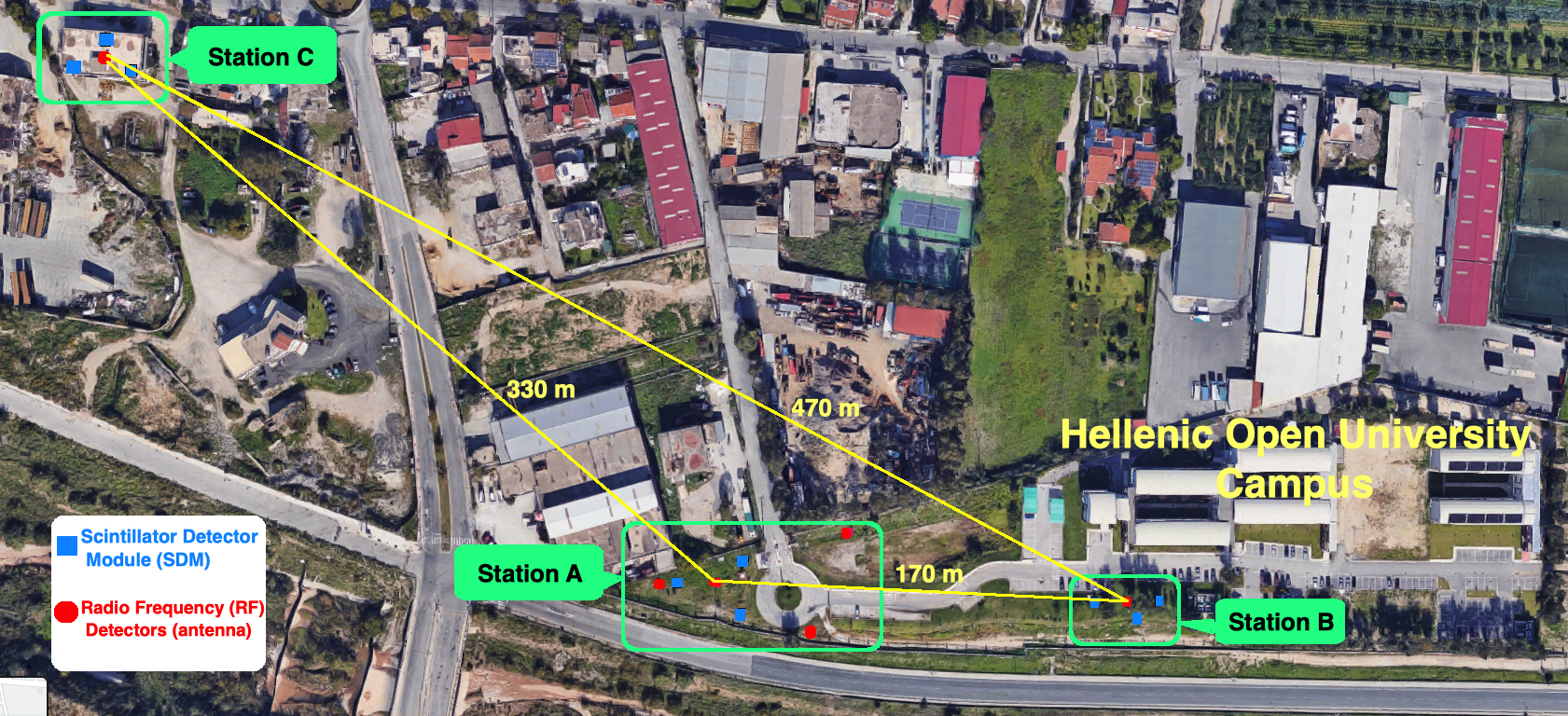}
\caption{The stations of the Astroneu array as well as their relative distances,  at the site of the Hellenic Open University campus. The SDMs are indicated with blue squares, while the RF detectors (antennas) with red circles.}
\label{fig1}
\end{figure}

\begin{figure}[h!]
\includegraphics[width=\textwidth]{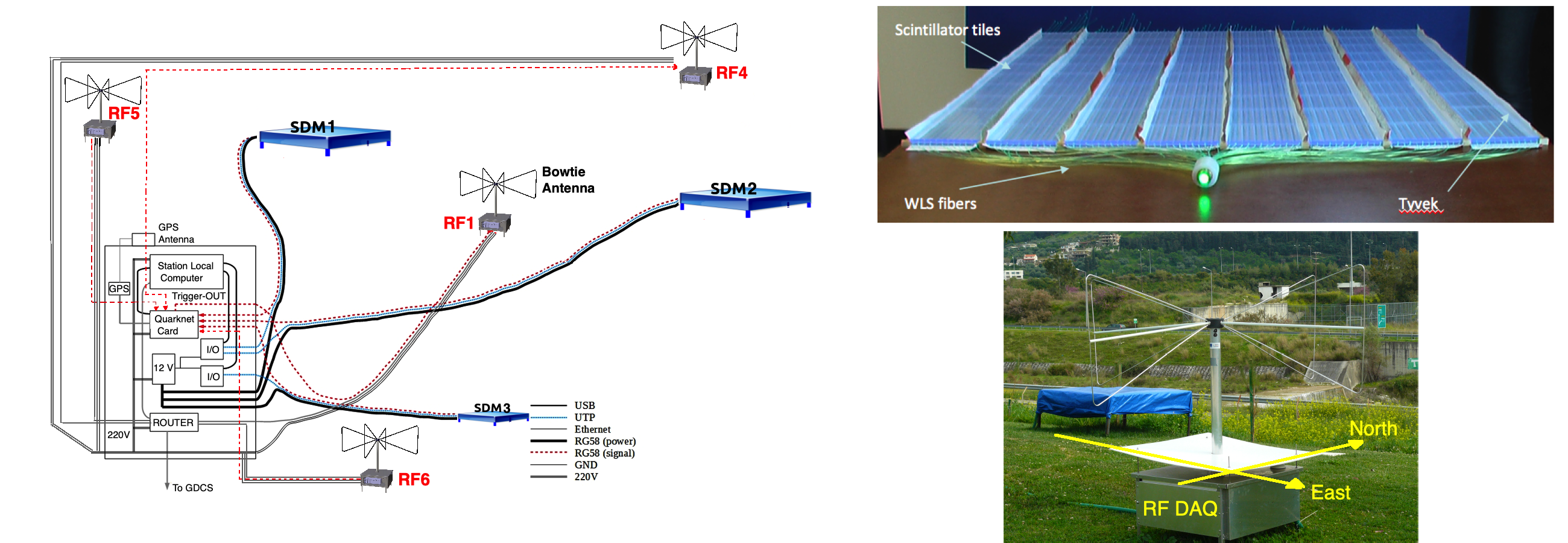}
\caption{{ \bf (Left)} The schematic illustration of the connections between the station A components, during the second operation period. {\bf (Right)} The photos depict the SDM during construction (top photo) and the RF bowtie antenna (bottom photo) structure.} 
\label{fig2}
\end{figure}

\paragraph*{} The Astroneu array \cite{5,6} is a hybrid EAS detection array operating inside the HOU campus. The location of the stations and their relative distances are depicted in Figure \ref{fig1}. During the first phase of operation (2014-2017), Astroneu composed of 3 autonomous stations (station-A, station-B and station-C) each one consisting of 3 large SDM and one RF detector (antenna). In the second operation period (2017-2022) three more RF detectors were installed in station-A as shown on the left side of Figure \ref{fig2}. The other two stations (B and C) were not changed. In station-A the positions of the four RF antennas were chosen in such a way that by combining three of them, triangles of different kinds (equilateral or amblygonal) and dimensions are formed. This offers the opportunity to study  the performance of different  geometrical layouts when only antennas are used to reconstruct the direction of the EAS axis, i.e. using the timing of  three RF signals and the positions of  three antennas.

The SDMs are made of two layers of plastic scintillation tiles each covering an area of approximately 1 m$^2$ (80 tiles per layer). The generated light by the scintillator is driven through wavelength shifting fibers (WLS) to a Photomultiplier tube (PMT) and transformed into a voltage signal. The three SDM PMTs pulses are received by a Quarknet board \cite{11} which measures the crossing time of the pulses waveforms with a predetermined amplitude threshold (which is set to 9.7 mV) with an accuracy of 1.25 ns. The instant of the first crossing is GPS tagged and defines  the pulse arrival time while the period of time that the waveform remains above the threshold (Time over Threshold—[ToT]) is used for the estimation of the pulse size. A detailed schematic representation of the connections between the Quarknet board and the SDMs of station-A is shown in Figure \ref{fig2} (right).

The RF detector is an antenna with two orthogonal bowtie shaped dipoles connected to a low noise amplifier (LNA), mounted on the top of the central support pole as shown in Figure \ref{fig2} (right). These types of antennas were constructed, calibrated and first used by the CODALEMA experiment \cite{7,12}. Their design has been fixed in order to be broadband (1-200 MHz), isotropic and sensitive to weak electromagnetic signals (as those emitted from EAS). The RF system is equipped with dedicated electronics and data acquisition (DAQ) system providing the opportunity for both self and external trigger operation. In the Astroneu array we use the latter option where the external trigger is provided by the Quarknet board upon the reception of  three SDM signals that exceed the voltage threshold of 9.7 mV. By receiving such a trigger the last 2560 sampled data (corresponding to a 2560 ns record) from both antenna's polarizations (in EW and NS direction) are digitized and stored. The GPS card of the RF system provides the appropriate time-stamps for the registered events. 
The correlation of the data recorded by the SDMs and the RF system is carried out offline by making use of the corresponding GPS time-tags. 
Following the same methodology,  data  from different stations of the telescope are combined.

\section{Data Sample \& Simulations}\label{sec2}
\subsection{Data Sample - RF event selection}\label{subsec2.1}
\paragraph*{} In this work the data sample was collected by the Astroneu's station-A (3SDM-4RF) over a period of four years (2017-2021) corresponding to an operating time of approximately 25,500 h. The initial sample composed of 480,000 EAS events detected and reconstructed by the 3 SDMs of the station with an energy threshold of 10 TeV. As expected for energies below 10$^{4}$ TeV the RF signals from EAS are very weak and therefore undetectable especially in areas with high levels of electromagnetic interference (coming mainly from manmade and stationary sources). Therefore the vast majority of these events, as expected, contain no detectable RF signal.

 For the 4$\times$480,000 recorded RF waveforms (from the 4 RF detectors of station-A) the offline RF event selection algorithm was deployed (see ref. \cite{8,9,10} for analytic description).  
The event selection algorithm for the RF data is a procedure consisting of 2 steps: the filtering process and the noise rejection algorithm. The filtering process, that uses a Tukey window filter and a Fast Fourier Transform (FFT), rejects  waveform frequencies outside the region 30-80 MHz (bellow 30 MHz
the ionospheric noise is enhanced, while above 80 MHz parasitic signals from  radio FM band are expected). Since the RF signals emitted by EAS are characterized by an intense peak located in a short time interval (transient) of approximately 30 ns, the filtered waveforms are checked for transients within the ``signal window'' of the 2560 ns buffer. The ``signal window'' is the time window inside the 2560 ns buffer where the  signal waveform should appear (according to the trigger signal) and ot is defined  between 1000 ns and 1500 ns. 
 Waveforms with mean power (in the ``signal window'') 6 times greater than the mean power of noise are retained to the next analysis stage. Of course transients from parasitic sources are also expected  especially in an area with powerful electromagnetic background. Therefore the second stage of the event selection algorithm is applied with the aim of rejecting these parasitic transients.   

The two waveforms (EW and NS polarizations) are digitized with a sampling rate of 1 GHz. For each waveform 2560 voltage values are stored in the array $V(i),~i=1,...,2560$. 
The first subroutine of the noise rejection algorithm checks the values of the rise time ($R_\mathrm{t}$) of the RF signal, defined as the time interval between two fixed values of the normalized cumulative function $C(i)$\footnote{$C(i)=\left(\sum_\mathrm{k=1000}^{1000+i}{V(k)^2}\right)\cdot\left(\sum_\mathrm{k=1000}^{1500}{V(k)^2}\right)^{-1}$, where $V(k)$ is the filtered RF signal. } of the signal. An update version of this subroutine was used in this work, compared to the one reported in \cite{8,9}, where the $R_\mathrm{t}$ is calculated between 25 \% and 65 \% of the cumulative function ($R_\mathrm{t}=C(65 \%)-C(25 \%)$) and events with $R_\mathrm{t}>$ 20 ns are rejected. This improves the noise rejection efficiency without significant loss of data as concluded from simulation samples. The second stage of the noise rejection algorithm involves the estimation of the degree of polarization 
(d.o.p.\footnote{$d.o.p.=\frac{\left(Q^2+U^2+V^2\right)^{-1/2}}{I}$, where Q, U, V, I are the Stokes parameters of the signal. For ideal linearly polarized signals $\mathrm{d.o.p} = 1$.}) 
of the RF transient. In contrast to parasitic signals the pulse emitted during EAS development is expected to be linear polarized. The threshold value for the d.o.p. was set to 0.85 (i.e. events with $\mathrm{d.o.p} < 0.85$ were rejected).

 The above criteria resulted in a final RF data sample of 460 events with signals in all four antennas of the station.      

\subsection{RF Simulation Package}\label{subsec2.2}
\paragraph*{} The simulation package is a four-step process. The first step involves the generation of high-energy EAS (10$^{17}$-2$\cdot$10$^{18}$ eV) using the MC simulation package CORSIKA \cite{13}. The QGSJET-II-04 \cite{14} and GEISHA \cite{15}  packages have been applied for high and low energy hadronic interactions, respectively, while the EGS4 Code \cite{16} was used for the electromagnetic interactions. In producing very high energy EAS the distributions of the primary particle, direction and spectral index were used according to recent measurements described in \cite{17,18,19}. Finally, the EAS core position was uniformly distributed in a radius of 400 meters around the center of the station. The MC sample size at this stage consisted of approximately 100000 events.    

\begin{figure}[h!]
\includegraphics[width=\textwidth]{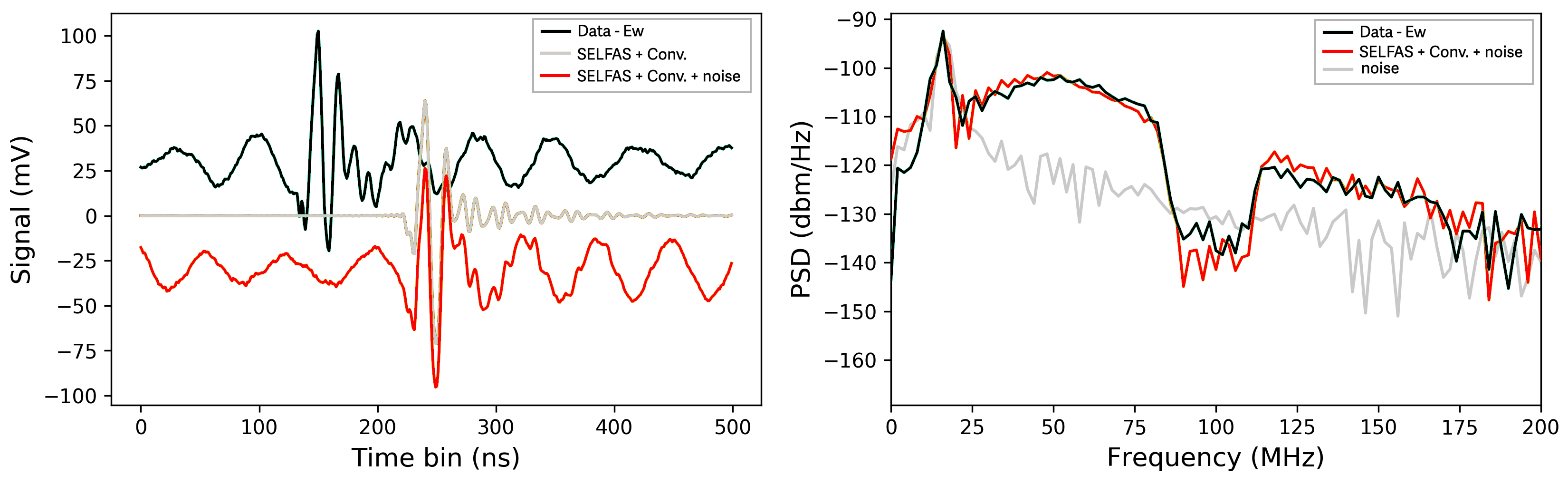}
\caption{{ Comparison at the waveform level between a real event and a simulated one with the same direction and shower core. \bf (Left)} The EW waveform (black line) as recorded by an antenna of station-A, the convolved SELFAS simulation (grey line) and the final simulated waveform as obtained after adding the noise (red line). Both data and simulated events are filtered in the frequency range 30-80 MHz. {\bf (Right)} The power spectrum densities (in dbm/Hz) for the recorded RF signal (black) in comparison with the simulated one (red line) and the noise (grey line).} 
\label{fig3}
\end{figure}

The second step in the simulation procedure involves the generation of the electric field produced by EAS, where the RF simulation code SELFAS \cite{20} was used (especially the updated version SELFAS3). Since SELFAS calculates electric field values only,  the third step is the convolution of the electric field  with the impulse response\footnote{The NEC package was used to simulate the RF system response as referred in \cite{10}.} of the RF system (antenna + LNA). In the next stage the simulated voltage signal is distorted by adding electromagnetic noise as measured in the area around the station for a period of one year. In the final stage, the event selection algorithm was imposed as described in Section \ref{subsec2.1}. The simulation sample after the described procedure was reduced to 30,500 events which was used for the rest of the analysis. In Figure \ref{fig3} (left) the EW waveform of a reconstructed event by an antenna of station-A is  shown (black line). In the same plot it is also shown the EW waveform (before and after the addition of noise) of a simulated event with almost the same direction and shower core as the real event\footnote{The esimation of the shower axis direction and core position is describe in the following sections.}. The corresponding power spectrum of these waveforms are depicted in Figure \ref{fig3} (right). In both cases the agreement is satisfactory. 

\section{ The effect of the RF antennas geometrical layout on the EAS direction resolution}\label{sec3}

The estimation of the EAS axis arrival direction (the zenith -$\theta$ and azimuth -$\phi$ angles) was implemented by using the arrival time of the RF signals, as well as the detector positions. The selected conventions are to define  East at $\phi=0^\circ$ and  North at $\phi=90^\circ$. For the reasons detailed in \cite{12}, the arrival time $t_{\mathrm{det}, k}$ in the $k^{th}$ RF detector is defined as the time of the maximum of the RF signal's envelope\footnote{The signal's enveloped is defined as $\tilde{V}=\sqrt{V^2(t)+\left({\cal{H}}V(t)\right)^2}$, where $V(t)$ is the signal and ${\cal{H}}V(t)$ its Hilbert transform.}. In first approximation the EAS axis direction ($\theta$, $\phi$) is estimated assuming plane wavefront for the RF pulse and a plane fit is implemented by minimizing the quantity

\begin{equation}
\chi^2=\sum_\mathrm{k=1}^{N}{\frac{\left(c(t_{\mathrm{det}, k}-t_0)-(A\cdot x_\mathrm{k}+B\cdot y_\mathrm{k})\right)^2}{\sigma_k^2}},
\end{equation}\label{eq1}
where $N$ is the number of the considered RF detectors, ($\it{x}_\mathrm{k}$, $\it{y}_\mathrm{k}$) the position coordinates of the $k^\mathrm{th}$ detector and $t_\mathrm{0}$ the arrival time at the origin of the coordinate system (0, 0). The quantity $\sigma_\mathrm{k}$ corresponds to the resolution in estimating the time in the $k^\mathrm{th}$ detector as calculated in \cite{9} varying from 8.4 to 8.9 ns for the 4 RF detectors of station-A. The zenith and azimuth angles of the EAS axis are determined using the equations 

\begin{equation}
\theta=\arcsin{\left(\sqrt{A^2+B^2}\right)}, \quad \phi=\arctan{\left(\frac{B}{A}\right)}.
\end{equation}\label{eq2}

\begin{figure}[h!]
\includegraphics[width=\textwidth]{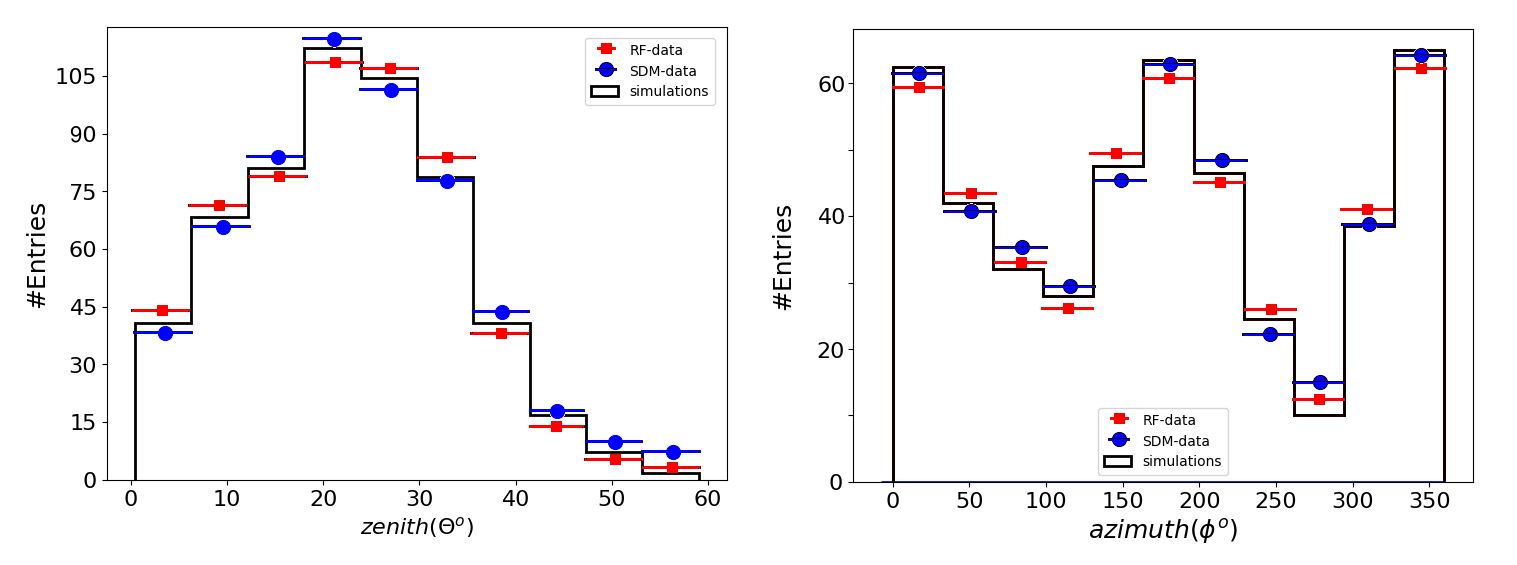}
\caption{{ \bf (Left)} The distribution of the reconstructed zenith angle  using the timing data from the 4 RF detectors (red squares), as well as the corresponding distribution using the timing data from the 3 SDM detectors (blue circles) of station-A. The histogram corresponds to the simulations predictions. {\bf (Right)} The same for the azimuth angle.} 
\label{fig4}
\end{figure}

\begin{figure}[h!]
\includegraphics[width=\textwidth]{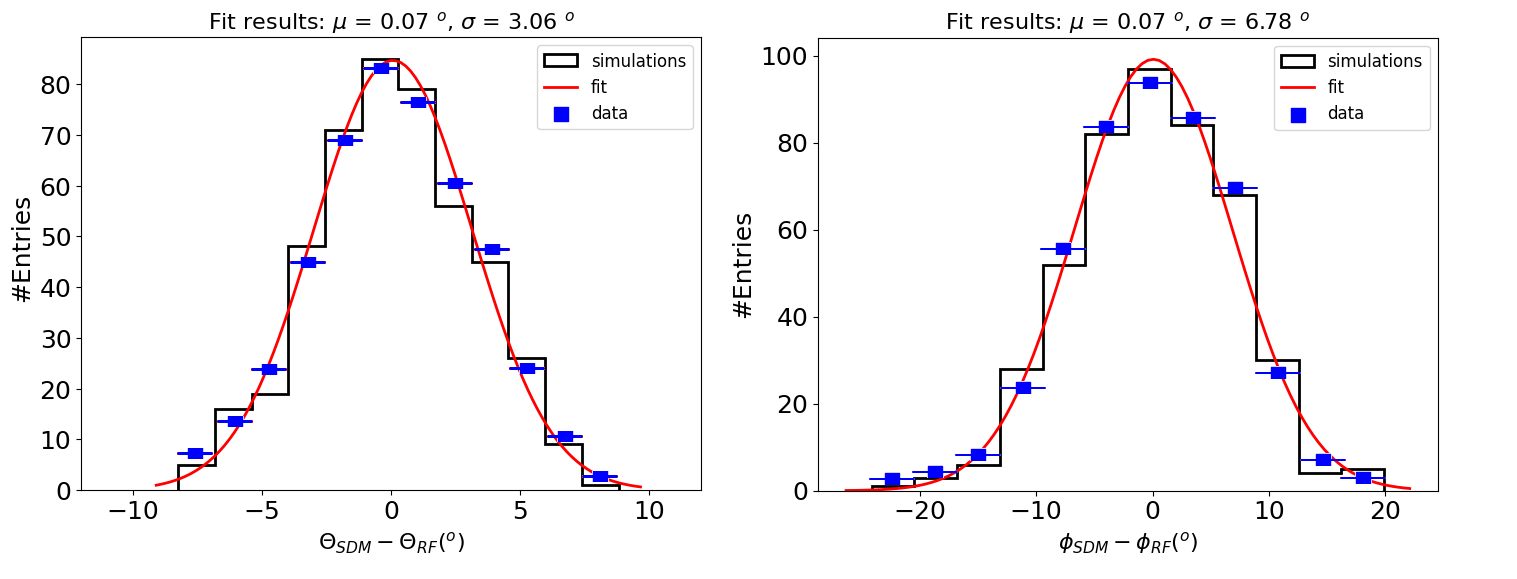}
\caption{{\bf (Left)} The distribution of the zenith angle difference $\Delta\theta=\theta_\mathrm{SDM}-\theta_\mathrm{RF}$ between the RF detector and the SDM detector measurements (blue points), compared to the simulation prediction (histogram). The red line corresponds to the Gaussian function that best fits the distribution. {\bf (Right)} Same for  the azimuth angle difference $\Delta\phi=\phi_\mathrm{SDM}-\phi_\mathrm{RF}$.} 
\label{fig5}
\end{figure}

In the first stage of the present analysis the EAS axis direction was reconstructed using timing data from  the 4 RF detectors only, as well as  timing data from the 3 SDM detectors only. The results were cross-correlated with the predictions of the MC simulations. Figure \ref{fig4} shows the zenith (left) and the azimuth angle (right) distributions as reconstructed from the RF system (red squares), as well as the corresponding distributions using the SDM (blue circles) system. The histogram corresponds to the MC predictions. 
Moreover Figure \ref{fig5} shows the differences of the zenith ($\Delta\theta=\theta_\mathrm{SDM}-\theta_\mathrm{RF}$, left) and  azimuth angle ($\Delta\phi=\phi_\mathrm{SDM}-\phi_\mathrm{RF}$, right) as derived from the reconstructions based on the RFs and SDMs data (blue points), compared to  simulations (histogram). Both distributions are well fitted (red line) with Gaussian functions of $\sigma_{\Delta\theta}=3.06^\circ\pm0.43^\circ$ and $\sigma_{\Delta\phi}=6.78^\circ\pm0.86^\circ$ for the zenith and azimuth angle differences respectively. 

\begin{figure}[h!]
\includegraphics[width=\textwidth]{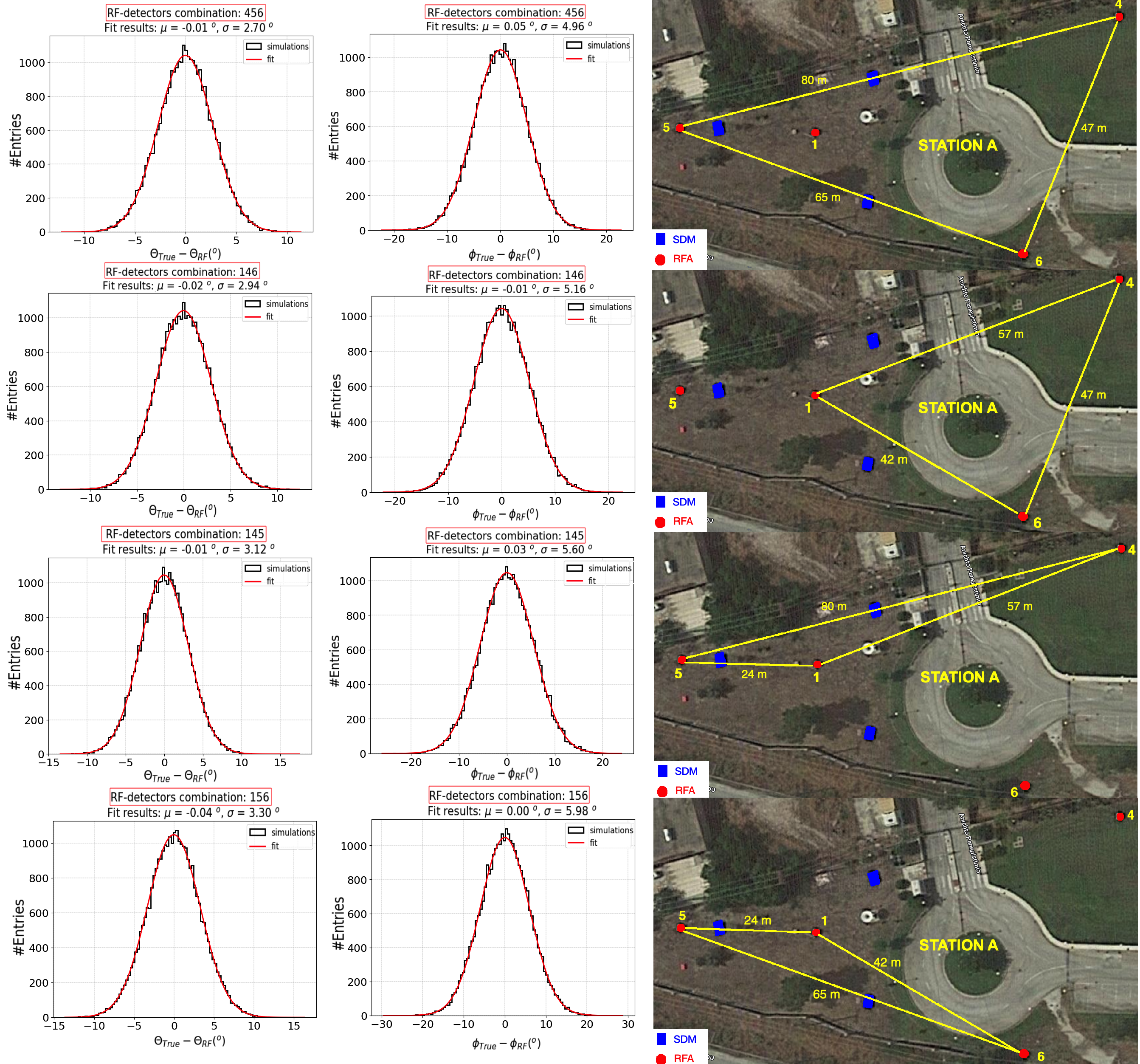}
\caption{{ \bf (Left)} The distributions of the difference between  true and  estimated EAS axis direction angles for the considered RF antenna geometries. {\bf (Right)} The corresponding antennas positions that constitute the four examined  geometries.} 
\label{fig6}
\end{figure}

In the next stage of the analysis, in order to correlate the effect of the station geometry to the resolution of the EAS axis reconstruction, four combinations of three RF detectors were used. As indicated in Figure \ref{fig6} (right), two of the four RF antenna combinations form an approximately isosceles triangle (456, 146) while the remaining two form an amblygonal triangle (145, 156). Between formations 456 and 146, the former exhibits greater distances separating the detectors. Likewise, the triangle formed by the antennas 145 has a larger diameter compared to the triangle of the antennas 156. The resolution of the described geometries in reconstructing the EAS axis direction was initially calculated for the simulation sample where the true direction is known (as used in the simulation's input file). In Figure \ref{fig6} (left) is presented the distributions of the difference between the true and the estimated (using the RF simulations sample), zenith ($\theta_\mathrm{true}-\theta_\mathrm{RF}$) and azimuth angles ($\phi_\mathrm{true}-\phi_\mathrm{RF}$). All distributions were fitted to Gaussian functions while fittings results are shown in Table \ref{tab1}. 
\begin{table}[h!] 
\caption{The results from fitting the Gaussian functions to the difference ($\theta_\mathrm{true}-\theta_\mathrm{RF}$), ($\phi_\mathrm{true}-\phi_\mathrm{RF}$) distributions for the considered RF-detectors formations.\label{tab1}}
\newcolumntype{C}{>{\centering\arraybackslash}X}
\begin{tabularx}{\textwidth}{CCCCC}
\toprule
 RF-detectors &  \textbf{456}	& \textbf{146}	& \textbf{145} & \textbf{156} \\
\midrule
$\mu_{(\theta_\mathrm{true}-\theta_\mathrm{RF})(^\circ)}$ & $-0.01\pm0.55$		& $-0.02\pm0.47$ & $-0.01\pm0.41$ & $-0.04\pm0.59$ \\
$\sigma_{(\theta_\mathrm{true}-\theta_\mathrm{RF})(^\circ)}$ & $2.70\pm0.34$		& $2.94\pm0.37$ & $3.12\pm0.41$ & $3.30\pm0.39$ \\
$\mu_{(\phi_\mathrm{true}-\phi_\mathrm{RF})(^\circ)}$ & $0.05\pm0.53$		& $-0.01\pm0.44$ & $0.03\pm0.61$ & $0.00\pm0.61$ \\
$\sigma_{(\phi_\mathrm{true}-\phi_\mathrm{RF})(^\circ)}$ & $4.96\pm0.64$		& $5.16\pm0.58$ & $5.60\pm0.59$ & $5.98\pm0.55$ \\
\bottomrule
\end{tabularx}
\end{table}
\unskip

\begin{figure}[h!]
\includegraphics[width=\textwidth]{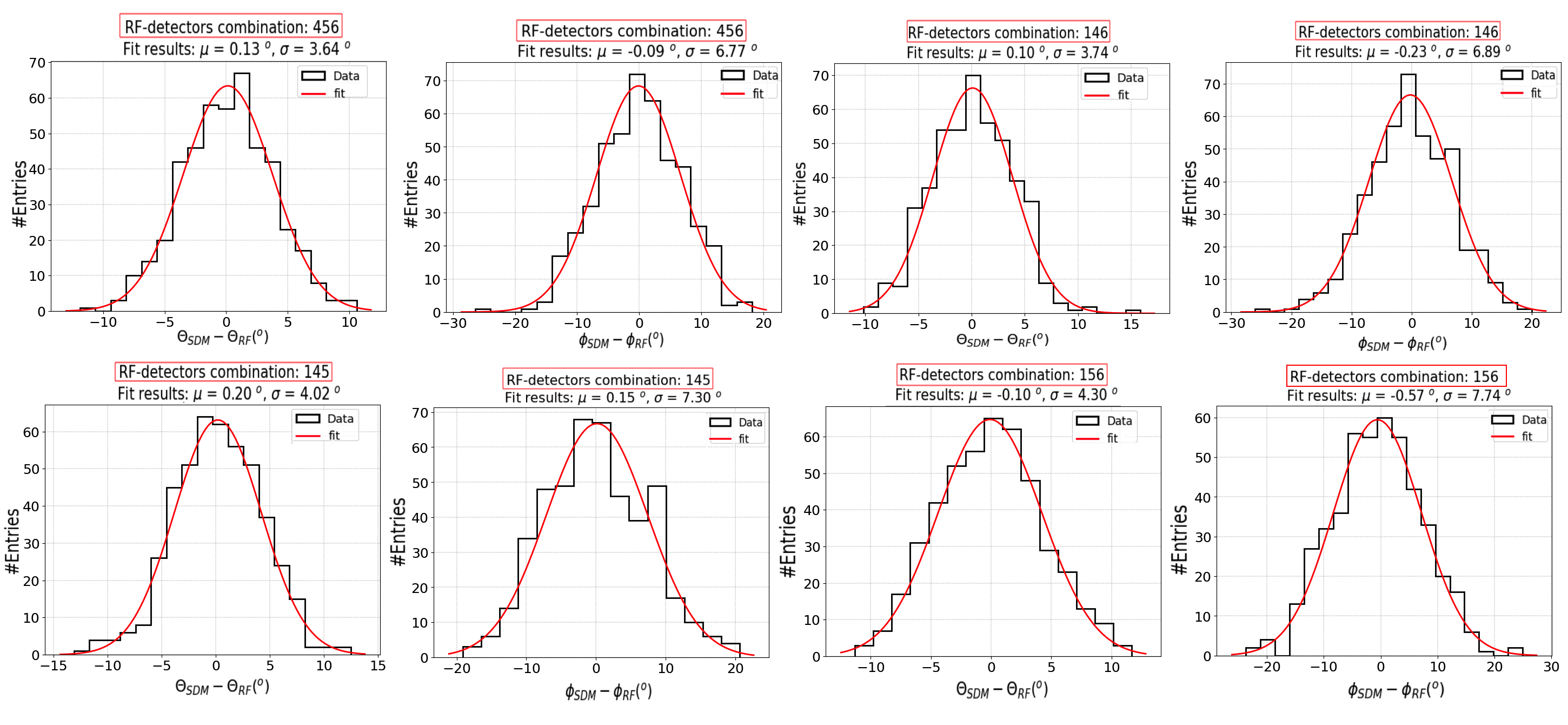}
\caption{{ \bf } Distributions of $\theta_\mathrm{SDM} - \theta_\mathrm{RF}$ and $\phi_\mathrm{SDM}-\phi_\mathrm{RF}$ (histograms) and the corresponding gaussian functions that best fit the data (red lines) for all four considered antenna formations.} 
\label{fig7}
\end{figure}

It is evident from the aforementioned results that among formations of the same shape, increasing the distance between detectors increases the accuracy of EAS axis direction reconstruction. Furthermore, as the shape of the triangle is converted from obtuse to (approximately) equilateral, the resolution in reconstructing the axis direction is improved. The next step of the analysis involves the correlation of the EAS axis directions as derived from measurements using the considered RF antenna formations and as extracted from the SDMs data independently. In Figure \ref{fig7} is shown the distribution of the difference between the reconstructed zenith (left) and azimuth (right) angle using the SDMs data and the RF data from the different antenna formations (histogram) while the red line corresponds to the fitted Gaussian function. The corresponding fitting results are represented in Table \ref{tab2}. In all formations the sigmas of the distributions are consistent with the individual resolutions of the SDM system\footnote{The zenith and azimuth resolution of the SDM system is $\sigma_{\Delta\theta}=2.40^\circ$, $\sigma_{\Delta\phi}= 4.60^\circ$ \cite{5}} and the corresponding RF formation (Table \ref{tab1}) which means that the sigmas are equal to the square root of the quadratic sum of the individual resolutions achieved by the two systems.

\begin{table}[h!] 
\caption{Gaussian fitting results for the difference ($\theta_\mathrm{SDM}-\theta_\mathrm{RF}$), ($\phi_\mathrm{SDM}-\phi_\mathrm{RF}$) distributions for the considered RF-detectors formations.\label{tab2}}
\newcolumntype{C}{>{\centering\arraybackslash}X}
\begin{tabularx}{\textwidth}{CCCCC}
\toprule
 RF-detectors &  \textbf{456}	& \textbf{146}	& \textbf{145} & \textbf{156} \\
\midrule
$\mu_{(\theta_\mathrm{SDM}-\theta_\mathrm{RF})(^\circ)}$ & $0.13\pm0.53$		& $0.10\pm0.67$ & $0.21\pm0.61$ & $-0.14\pm0.69$ \\
$\sigma_{(\theta_\mathrm{SDM}-\theta_\mathrm{RF})(^\circ)}$ & $3.64\pm0.84$		& $3.74\pm0.87$ & $4.02\pm0.91$ & $4.30\pm0.89$ \\
$\mu_{(\phi_\mathrm{SDM}-\phi_\mathrm{RF})(^\circ)}$ & $-0.059\pm0.53$		& $-0.23\pm0.54$ & $0.15\pm0.60$ & $-0.57\pm0.81$ \\
$\sigma_{(\phi_\mathrm{SDM}-\phi_\mathrm{RF})(^\circ)}$ & $6.77\pm0.94$		& $6.89\pm0.88$ & $7.30\pm0.89$ & $7.74\pm0.95$ \\
\bottomrule
\end{tabularx}
\end{table}
\unskip

\section{EAS core reconstruction}\label{sec4}
\begin{figure}[h!]
\includegraphics[width=\textwidth]{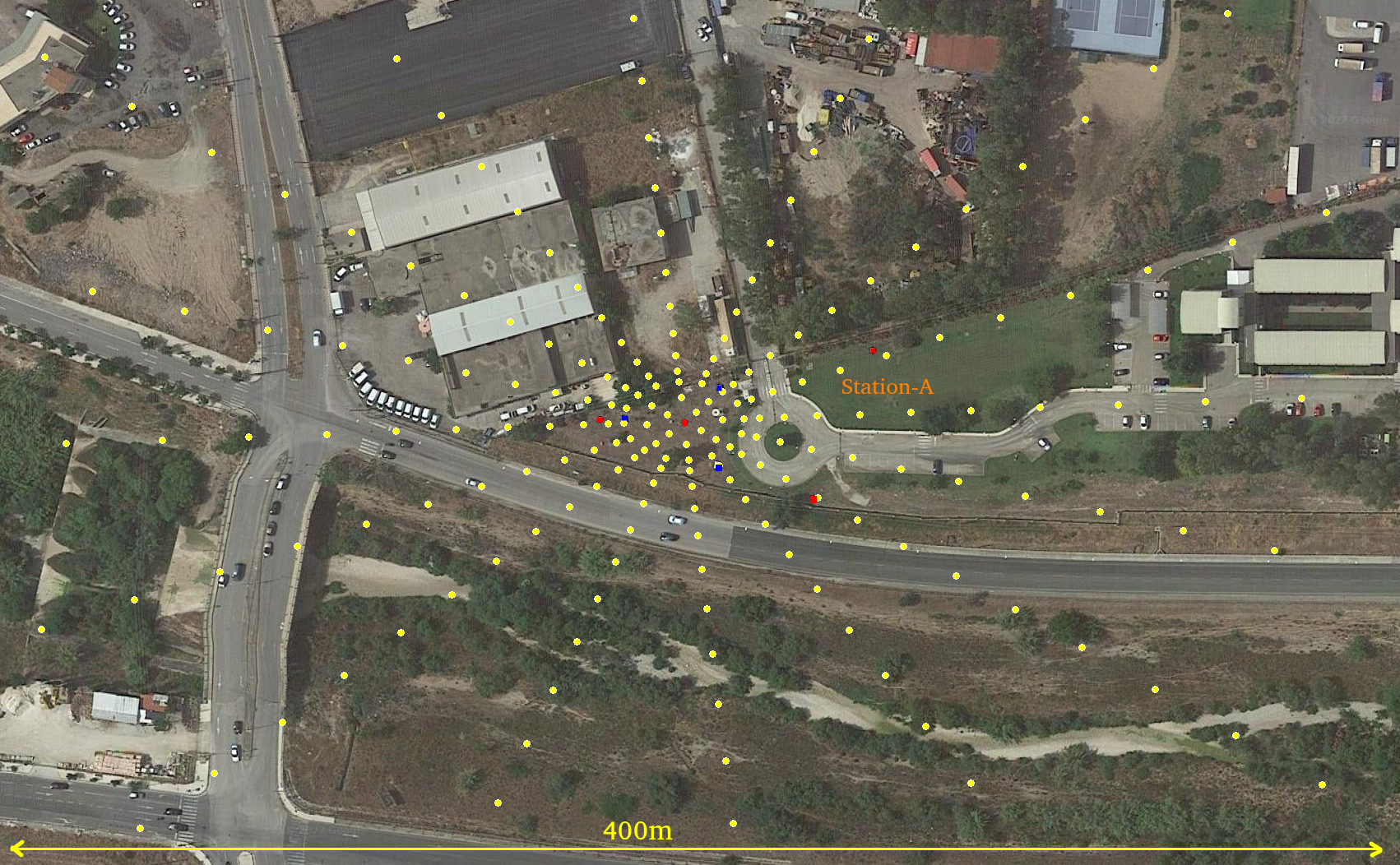}
\caption{The layout of the points (yellow circles) used for the electric field calculation at ground level from the set of 60 simulations. The positions of the RF and SDM detectors are marked with the red and blue points respectively.} 
\label{fig8}
\end{figure}

\paragraph*{} The term "shower core" is commonly used for the intersection point of the EAS axis with the ground level. In this study we present a new method of estimating the EAS core by correlating the electric field map as measured on the ground with the expected field values estimated using MC simulations. 
For each event identified as of cosmic origin a set of 60 simulations was prepared with the specifications described in Section \ref{subsec2.2}. In these MC events the electric field at the ground level was calculated at 220 points spread in a radius of approximately 250 m around the center of the station-A as depicted in Figure \ref{fig8}. These  positions were chosen with the criterion that the electric field can be calculated also at the intermediate points using linear interpolation. For the production of the simulations the primary direction (zenith and azimuth angle) was fixed to the reconstructed direction using the RF data, while the primary energy was set arbitrary to 10$^{18}$ eV \footnote{Such events induce RF signals considerably above the electromagnetic background.}. The core position was placed at the center of station-A  and for the primary particle we used protons (p) and iron nuclei (Fe) since these are expected to be the main CR candidates at ultra high energies. Out of 60 simulations, 40 corresponded to proton primary and 20 to iron nuclei enclosing a sensible number of $X_\mathrm{max}$ values in accordance with the fluctuations that appear between EAS of the same energy\footnote{The larger number of simulations for the proton case is justified by the fact that its interaction cross section with air nuclei leads to larger $X_\mathrm{max}$ fluctuations.}. 
Averaging  the simulated electric fields of these 60 showers, the map of the electric field on the ground is obtained when the shower core $(x_\mathrm{cr},y_\mathrm{cr})$ is located at the center of station-A i.e. $(x_\mathrm{cr},y_\mathrm{cr})=(0,0)$. By moving the field map within a circular disk of radius 200 m\footnote{In order to increase the number of iterations and reduce the computational time.} around the center of the station, a new field map could be obtained for a different $(x_\mathrm{cr},y_\mathrm{cr})$ position. In order to estimate the shower core position from the RF data a  fitting procedure was employed where the measurements of the 4 antennas $V_\mathrm{det}(x_\mathrm{k},y_\mathrm{k}),~k=1,...,4$, were compared with a large number\footnote{The fields maps were produced with a granularity of 1 m in both $x$, $y$ directions.} of fields maps each one corresponding to different shower core position $(x_\mathrm{cr},y_\mathrm{cr})$. Then the estimation of the shower core position from the data was obtained by searching the minimum value of the quantity 

\begin{equation}
\chi^2(x_\mathrm{cr},y_\mathrm{cr})=\sum_\mathrm{k=1}^{4}{\left(V^{k}_\mathrm{det}-V^{k}_\mathrm{sim}(x_\mathrm{cr},y_\mathrm{cr})\right)^2/\sigma_\mathrm{k}^2},
\label{eq3}
\end{equation} 
where $V^{k}_\mathrm{det}$ is the measured pulse height of the k$^{th}$ RF antenna and $V^{k}_\mathrm{sim}(x_\mathrm{cr},y_\mathrm{cr})$ is the expected pulse height at the position of the k$^{th}$ RF antenna according to field map with shower core at $(x_\mathrm{cr},y_\mathrm{cr})$. Since the energy of the primary particle in the MC sample was fixed to $10^{18}~\mathrm{eV}$ and taking into account that the electric field and consequently the RF signal is proportional to the primary energy, a scaling factor $a$ was applied to $V^{k}_\mathrm{sim}(x_\mathrm{cr},y_\mathrm{cr})$. The scaling factor $a$ was estimated as $a=(1/4)\cdot\sum_{1}^{4}{\left(V^{k}_\mathrm{det}/V^{k}_\mathrm{sim}(x_\mathrm{cr},y_\mathrm{cr})\right)}$ i.e. the mean value of the scale factors between the 4 RF antennas. In this way we eliminate the effect of the shower energy to the strength of the shower signal and retain only the effect of the attenuation of the signal due to the distance to the shower axis. In future studies this scaling factor can be incorporated to the $\chi^2$ fitting procedure in order to estimate the energy of the shower.   
The term $\sigma_\mathrm{k}$ corresponds to the background RF interference as measured in the RF detectors positions, while the EAS core position (X$_\mathrm{cr}$, Y$_\mathrm{cr}$) corresponds to the values $(x_\mathrm{cr},y_\mathrm{cr})$ that minimize the $\chi^2$ value of equation \ref{eq3}. 
   
\begin{figure}[h!]
\includegraphics[width=\textwidth]{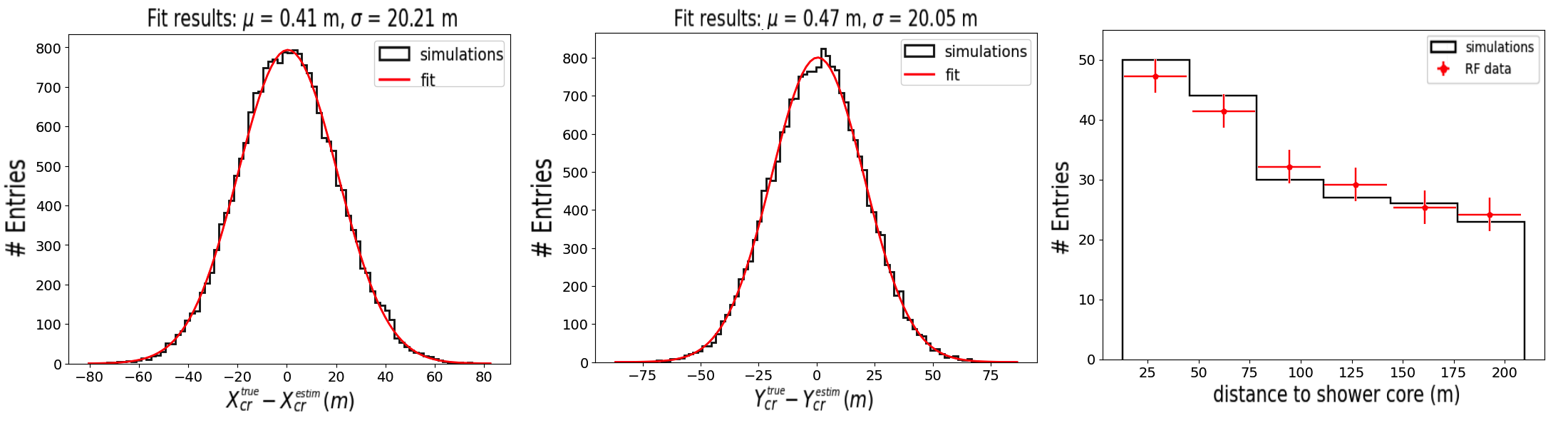}
\caption{ {\bf (Left)} The distribution of the difference between the true ($X^\mathrm{true}_\mathrm{cr}$) and the estimated ($X^\mathrm{estim}_\mathrm{cr}$) X-coordinate of the core position (histogram) for the simulation sample. The red curve represents the Gaussian function that fits the distribution.  {\bf (Middle)} Similar for the Y-coordinate of the core position. {\bf (Right)} The distribution of the EAS core distance  as derived from the RF data (red points) and simulations (histogram).} 
\label{fig9}
\end{figure}

The described method for the EAS core estimation was tested using the simulation sample (that consisted of 30500 events as described in Section \ref{subsec2.2}) where the true core position is known. In order to reduce significantly the required computational time\footnote{For each simulated shower 60 additional simulations are produced.}, only  events with true core position  located within a radius of 200 meters around the center of the station-A were selected (these corresponded to 20500 events). For each event  the core position was estimated (using the described method) and compared with the true one. The left plot of Figure \ref{fig9} represents the distribution of the difference between the true $X^\mathrm{true}_\mathrm{cr}$ and the estimated $X^\mathrm{estim}_\mathrm{cr}$   coordinate of the core position, while the corresponding distribution for the Y coordinate is shown on the middle plot of Figure \ref{fig9}. A Gaussian fit is imposed on both distributions with the resulting mean value close to zero and  standard deviation approximately 20 m for both coordinates. Subsequently, the method was applied to the collected RF data for the full sample of 460 events. Again only events whose core position were at distance less or equal to 200 m from the center of the station-A were selected. The right plot of Figure \ref{fig9} shows the distribution of these EAS core distances  for the RF data (denoted with the red points) in comparison with the simulation prediction (histogram). As no other method for determining the core position is available, the efficiency of the method was further tested by calculating the charge excess to geomagnetic ratio (as presented in Section \ref{sec5} / below) which strongly depends on the core position.

\section{Charge excess to Geomagnetic Ratio}\label{sec5}

\paragraph*{} As already mentioned in Section \ref{intro}, the directions of the electric field vectors associated with the two main mechanisms (Geomagnetic and Charge Excess) of the RF emission are different. The electric field ($\vec{E}_G$) related to the Geomagnetic mechanism is in the direction $\hat{n}_s\times\hat{n}_B$ ($\hat{n}_s$, $\hat{n}_B$ are the unit vectors in the directions of the EAS axis and the geomagnetic field, respectively) while the Charge Excess component ($\vec{E}_C$) is directed from the point of observation towards  the EAS axis. The measured EW and NS components of the EAS RF transient ($E_\mathrm{EW}$, $E_\mathrm{NS}$) can be expressed in terms of the projections of the two contributions in the ground plane ($E^\mathrm{pr}_\mathrm{G}$, $E^\mathrm{pr}_\mathrm{C}$) according to the formula 

\begin{equation}
E_\mathrm{EW}=E^\mathrm{pr}_\mathrm{G}\cos{\psi_\mathrm{G}}+E^\mathrm{pr}_\mathrm{C}\cos{\psi_\mathrm{C}}, \:\:\: E_\mathrm{NS}=E^\mathrm{pr}_\mathrm{G}\sin{\psi_\mathrm{G}}+E^\mathrm{pr}_\mathrm{C}\sin{\psi_\mathrm{C}},
\label{eq4}
\end{equation}
where $\psi_\mathrm{G}$, $\psi_\mathrm{C}$ are the angles formed between the EW direction and the projected Geomagnetic ($E^\mathrm{pr}_\mathrm{G}$) and Charge Excess ($E^\mathrm{pr}_\mathrm{C}$) electric fields respectively as depicted in  Figure \ref{figA1}. The $\sin{\psi_\mathrm{G}}$ and $\cos{\psi_\mathrm{G}}$ can be expressed in terms of the EAS axis ($\theta$, $\phi$) and the Geomagnetic field ($\theta_\mathrm{B}$, $\phi_\mathrm{B}$) directions while the $\sin{\psi_\mathrm{C}}$ and $\cos{\psi_\mathrm{C}}$ in terms of the core (X$_\mathrm{cr}$, Y$_\mathrm{cr}$) and the RF detector (x$_\mathrm{k}$, y$_k$) coordinates as discussed in Appendix \ref{A1}.

The contribution of each mechanism to the measured EAS RF signal can be quantified by the Charge Excess to Geomagnetic Ratio (C$_\mathrm{GRT}$) defined by the relation 

\begin{equation}
C_\mathrm{GRT}=\frac{E^\mathrm{pr}_\mathrm{C}}{\frac{E^\mathrm{pr}_\mathrm{G}}{\sin{\alpha}}}=\frac{E^\mathrm{pr}_\mathrm{C}}{E^\mathrm{pr}_\mathrm{G}}\cdot\sin{\alpha},
\label{eq5}
\end{equation} 
where $\alpha$ (geomagnetic angle) represents the angle between the EAS axis ($\hat{n}_s$) and the Geomagnetic field direction ($\hat{n}_B$). Since the Geomagnetic component is proportional to $\sin{\alpha}$ the term $E^\mathrm{pr}_\mathrm{G}/\sin{\alpha}$ express the relative strength of the mechanism, excluding C$_\mathrm{GRT}$ large values due to small geomagnetic angles\footnote{In this way, the C$_\mathrm{GRT}$ becomes independent from the azimuth angle of the EAS axis.}. 
The polarization angle ($\phi_p$)\footnote{The angle formed by the semi major axis of the polarization ellipse with the EW direction, $\tan{\phi_p}=\frac{E_\mathrm{NS}}{E_\mathrm{EW}}$} of the RF signal can be also expressed in terms of the C$_\mathrm{GRT}$ as derived combining equations \ref{eq4}, \ref{eq5}

\begin{equation}
\tan{\phi_p}=\frac{\sin{\psi_\mathrm{G}}+ C_\mathrm{GRT}\left(\sin{\alpha}\sin{\psi_\mathrm{C}}\right)^{-1}}{\cos{\psi_\mathrm{G}}+ C_\mathrm{GRT}\left(\sin{\alpha}\cos{\psi_\mathrm{C}}\right)^{-1}}.
\label{eq6}
\end{equation} 

On the other hand, using the recorded EW and NS waveforms the polarization angle  $\phi_p$ can  be estimated from the data:

\begin{equation}
\tan{\phi_p}=\frac{E_\mathrm{NS}}{E_\mathrm{EW}}.
\label{eq7}
\end{equation} 

Consequently the C$_\mathrm{GRT}$ can be estimated by these two expressions of the polarization angle assuming that the direction of the shower as well as the shower core is known. 

As demonstrated by simulation studies \cite{22} the $C_\mathrm{GRT}$ depends strongly on the opening angle from the point of the EAS maximum ($X_\mathrm{max}$) to the point of observation. In particular the $C_\mathrm{GRT}$ is expected to increase as the observation angle increases. Large opening angles correspond to large distances from EAS core for almost vertical EAS axis. For inclined EAS axis the point of the EAS maximum is further away from the observation point (related to a vertical one) which means that the same distance from the EAS core corresponds to a smaller opening angle and consequently to smaller $C_\mathrm{GRT}$ value. We can therefore conclude that the $C_\mathrm{GRT}$ value increases with increasing distance from EAS core and decreases with increasing EAS zenith  angle. These dependences of the $C_\mathrm{GRT}$ were studied in the present work, using both the experimental RF data as well as the simulation sample. 

Figure \ref{fig10} (left) shows the C$_\mathrm{GRT}$ variation for increasing distance from the shower core ($d_\mathrm{cr}$) considering four different bins of the zenith angle (from 0$^\circ$ to 60$^\circ$). The results from the RF data are represented with  red points. The black curve corresponds to the simulation sample. Similarly Figure \ref{fig10} (right) represents the C$_\mathrm{GRT}$ variation for increasing EAS axis zenith angle ($\theta$) considering four different distance ranges (from 0 m to 200 m). RF data are presented with red points while the black curve corresponds to the  simulation sample. In all cases the  C$_\mathrm{GRT}$ values are in agreement with those expected from simulations and verified in previous studies \cite{23,24}. A synopsis of the estimated C$_\mathrm{GRT}$ values is shown in Table \ref{tab3}.  

\begin{table}[h!] 
\caption{The summary of the estimated C$_\mathrm{GRT}$ values for different EAS zenith angles and distances from EAS core position.\label{tab3}}
\newcolumntype{C}{>{\centering\arraybackslash}X}
\begin{tabularx}{\textwidth}{CCCCC}
\toprule
 &$d\in[0,50]~m$&$d\in(50,100]~m$&$d\in(100,150]~m$&$d\in(150,200]~m$\\
\midrule
$\theta\in[0,15^\circ]$ &8.10\%&13.15\%&17.14\%&19.23\%\\
$\theta\in(15,30^\circ]$ &6.96\% &10.76\% &12.50\% &14.92\% \\
$\theta\in(30,45^\circ]$ &5.16\% &7.08\% &8.74\% &10.76\% \\
$\theta\in(45,60^\circ]$ &4.13\% &6.56\% &8.62\% &10.45\% \\
\bottomrule
\end{tabularx}
\end{table}
\unskip

\begin{figure}[h!]
\includegraphics[width=\textwidth]{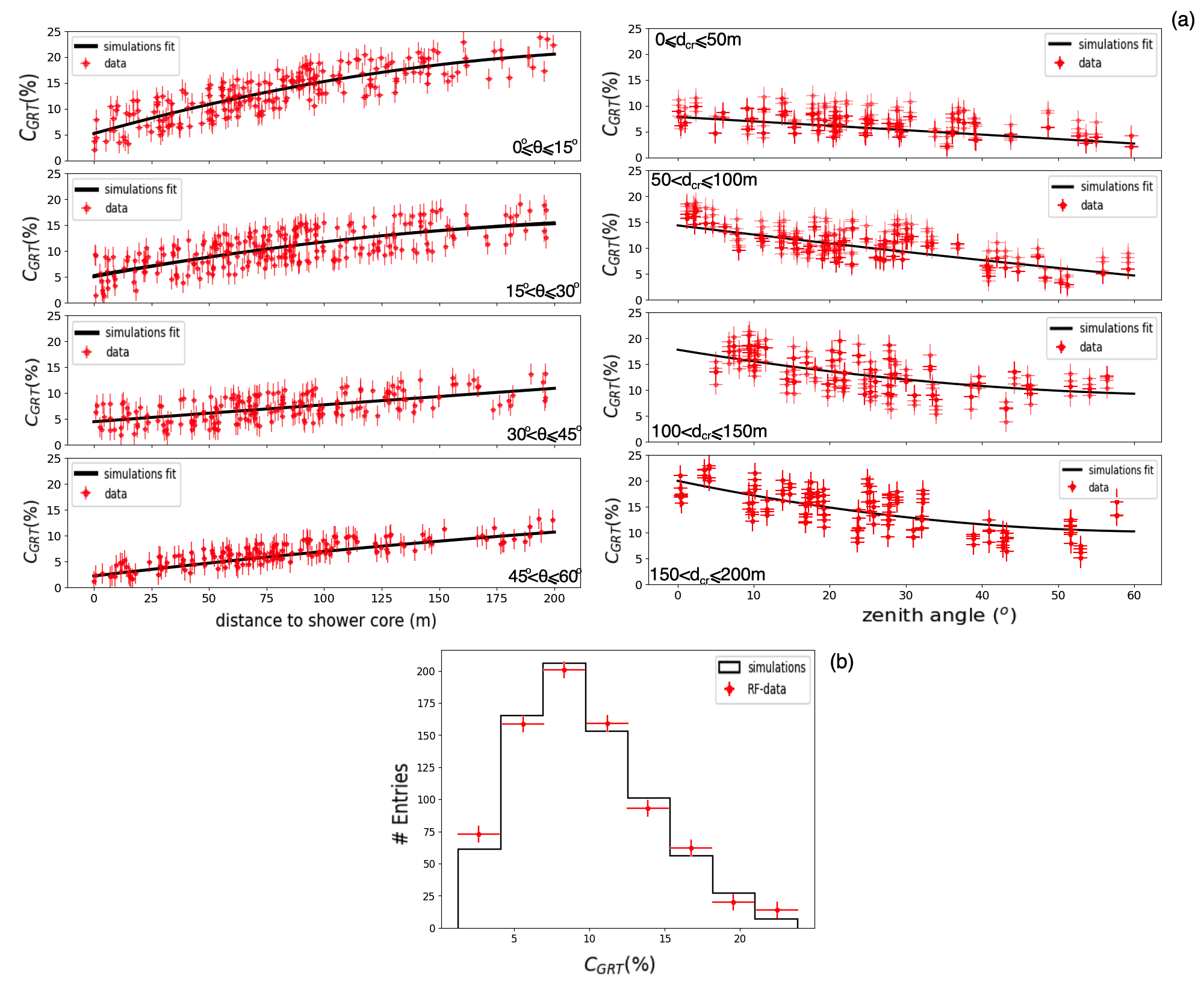}
\caption{{\bf (Upper-Left)} The C$_\mathrm{GRT}$ variation as a function of the distance to shower core for four different zenith angle ranges. The red points correspond to the RF data while the black curve to the simulations. {\bf (Upper-Right)} The C$_\mathrm{GRT}$ variation as a function of the zenith angle for four different core distance ranges. The red points correspond to the RF data while the black curve to the simulations. {\bf (Bottom)} The distribution of  C$_\mathrm{GRT}$ as estimated from the RF-data (red points) and simulations (black histogram). } 
\label{fig10}
\end{figure}
Figure \ref{fig10} (bottom) shows the distribution of the C$_\mathrm{GRT}$ values as reconstructed from simulations (black histogram) and data (red points). The two distributions are in good agreement. 
Since the values of the C$_\mathrm{GRT}$ are highly depended on the EAS core position, the apparent agreement of the C$_\mathrm{GRT}$ values presented in this study compared to previous ones and simulations is a strong evidence that the EAS core  reconstruction method is efficient.  

\section{Conclusions}\label{sec6}
\paragraph*{} The Astroneu array is small scale hybrid array operating in strong electromagnetic noise environment.
In this paper, we exploited further the RF signals captured by the array after proper filtering and noise rejection. 
For this purpose, data from various RF
antennas as well as large simulation samples were analyzed. The arrival times of the RF signals were used to estimate the EAS axis direction and it was found that the RF system measurements are in very good agreement with the particle detector measurements as well as with  simulation predictions. It was also shown that the reconstructed EAS direction resolution (around 3 degrees in zenith and 5.5 degrees in azimuth) improves for  equilateral geometrical layouts and larger distances between the detectors, a feature that is also known for particle detectors. A new method for the reconstruction of the EAS core using the expected field map of the electric field and the detailed response of the RF antennas was also presented. The reconstruction resolution was found to be approximately 20 m in both x and y directions for showers within an area of  radius 200 m around station-A of the Astroneu array.
The estimated shower core position was also used for the measurement of the relative strength between the two main mechanisms of radio emission by high energy showers i.e. the Charge Excess to Geomagnetic ratio, $C_\mathrm{GRT}$.  The estimated $C_\mathrm{GRT}$ values varies from 3\% for large zenith angles and small core distances up to 24\% for roughly vertical showers at large distances (>150 m). These measurements are in very good agreement with the expectations as well as other experimental  measurements. In the planned Astroneu array extension (with more RF and SDM detectors), $C_\mathrm{GRT}$ measurements will be used supplementary to the existing noise rejection algorithms aiming for the development of a self trigger operation mode. Finally, since the Charge Excess mechanism is the main contribution for RF emission in dense media, the analysis of $C_\mathrm{GRT}$ can also be used as a powerful tool for distinguishing  EAS propagating on different media. 

\section*{Acknowledgments}
\paragraph*{} This research was funded by the Hellenic Open University Grant No. $\Phi$K 228: ``Development of technological applications and experimental methods in Particle and Astroparticle Physics''

\newpage
\appendix
\section{EAS geometry in the ground plane}\label{A1}
\begin{figure}[h!]
\includegraphics[width=\textwidth]{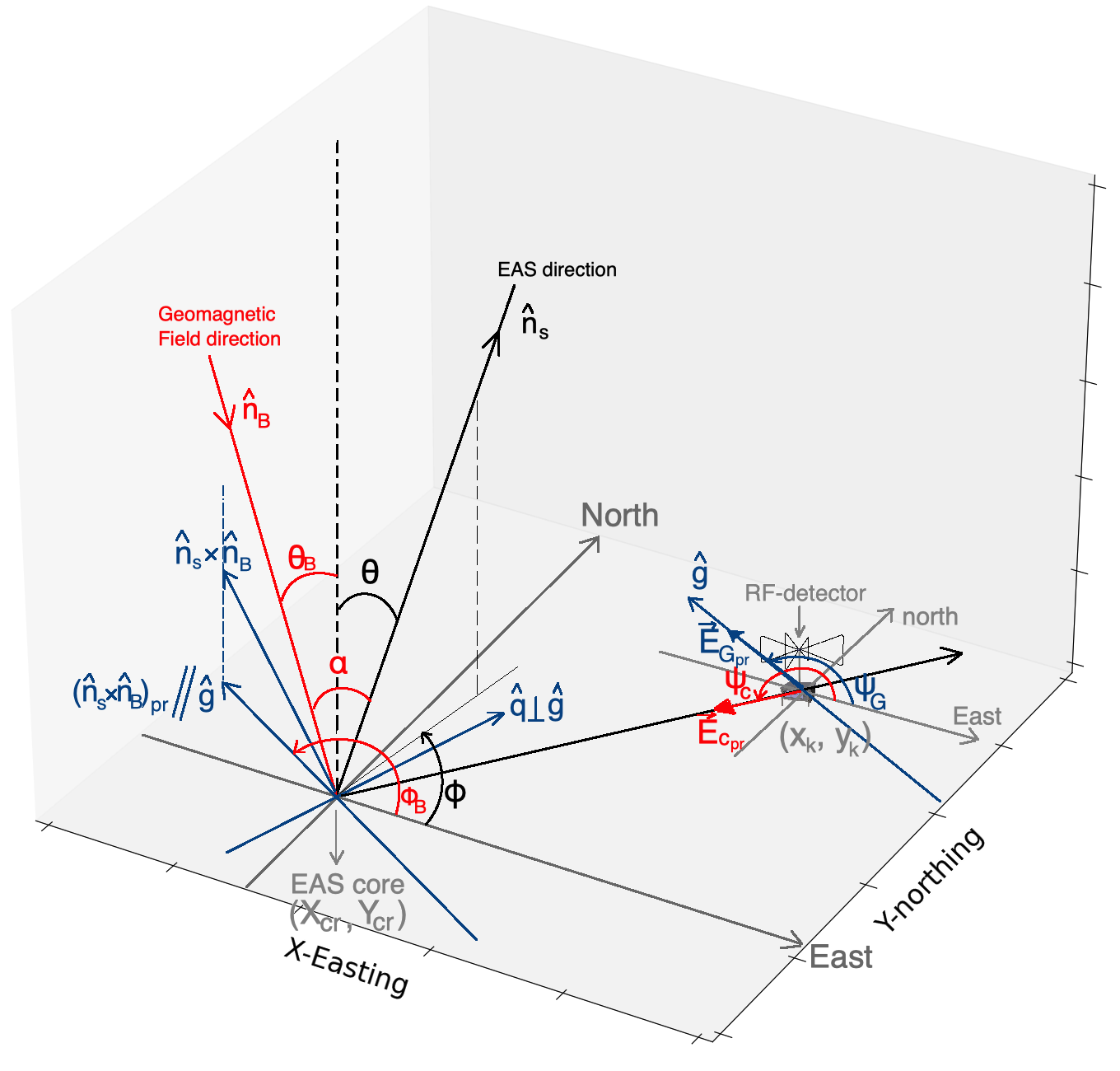}
\caption{An incoming EAS (with zenith and azimuth angles $\theta$, $\phi$ respectively) in the direction $\hat{n}_\mathrm{s}$ with respect to the direction of the geomagnetic field $\hat{n}_\mathrm{B}$ (with zenith and azimuth angles $\theta_B$, $\phi_B$ respectively). The angle formed by the EAS axis and the Geomagnetic field is denoted with $\alpha$. The projection of the vector $\hat{n}_\mathrm{s}\times\hat{n}_\mathrm{B}$ in the ground plane is in the direction of the g-axis. The q-axis is orthogonal to the g-axis. $\vec{E}_\mathrm{G_\mathrm{pr}}$ and $\vec{E}_\mathrm{C_\mathrm{pr}}$ denote the ground plane projections of the Geomagnetic and the Charge Excess electric field components respectively. The angle formed by the $\vec{E}_\mathrm{G_\mathrm{pr}}$ and the East direction is denoted as $\psi_\mathrm{G}$ while the corresponding angle of the $\vec{E}_\mathrm{C_\mathrm{pr}}$ as $\psi_\mathrm{ch}$. ($X_\mathrm{cr},Y_\mathrm{cr}$) and ($x_\mathrm{k},y_\mathrm{k}$) are the positions of the EAS core and the RF detector respectively.} 
\label{figA1}
\end{figure}

\paragraph*{} In this section we describe the formulas for calculating the angles $\psi_\mathrm{G}$ and  $\psi_\mathrm{C}$ of the electric fields $\vec{E}_\mathrm{G_\mathrm{pr}}$ and $\vec{E}_\mathrm{C_\mathrm{pr}}$  with the EW direction as depicted in Figure \ref{figA1}.
Both angles are defined to be zero at East and positive towards North. The $\sin{\psi_\mathrm{G}}$ and $\cos{\psi_\mathrm{G}}$ can be determined from the ground plane projection of the cross product of the unit vector along the EAS axis direction ($\hat{n}_\mathrm{s}$) and the unit vector in the direction of the Geomagnetic field ($\hat{n}_\mathrm{B}$) as given by the formulas 
\begin{equation}
\cos{\psi_\mathrm{G}}=\left(\hat{n}_\mathrm{S}\times\hat{n}_\mathrm{B}\right)_\mathrm{EW}=\sin{\phi}\sin{\theta}\cos{\theta_\mathrm{B}}-\sin{\phi_\mathrm{B}}\sin{\theta_\mathrm{B}}\cos{\theta} 
\label{eqA1}
\end{equation} 
\begin{equation}
\sin{\psi_\mathrm{G}}=\left(\hat{n}_\mathrm{S}\times\hat{n}_\mathrm{B}\right)_\mathrm{NS}=\cos{\theta}\cos{\phi_\mathrm{B}}\sin{\theta_\mathrm{B}}-\cos{\phi}\sin{\theta}\cos{\theta_\mathrm{B}}
\label{eqA2}
\end{equation} 
where $\theta$, $\phi$ are the zenith and azimuth angles, respectively, of the EAS axis direction and $\theta_\mathrm{B}$, $\phi_\mathrm{B}$ the corresponding angles of the Geomagnetic field direction (in the  Astroneu site $\theta_\mathrm{B}=34.67^\circ$ and $\phi_\mathrm{B}=273.54^\circ$). 
The $\sin{\psi_\mathrm{C}}$ and $\cos{\psi_\mathrm{C}}$ can be derived from the position of the EAS core ($X_\mathrm{cr},Y_\mathrm{cr}$) in the ground plane with respect to the RF detector position ($x_\mathrm{k},y_\mathrm{k}$) as shown in Figure \ref{figA1}
\begin{equation}
\sin{\psi_\mathrm{C}}=\frac{Y_\mathrm{cr}-y_\mathrm{k}}{d_\mathrm{cr}}, \:\: \cos{\psi_\mathrm{C}}=\frac{X_\mathrm{cr}-x_\mathrm{k}}{d_\mathrm{cr}}, \:\: d_\mathrm{cr}=\sqrt{(X_\mathrm{cr}-x_\mathrm{k})^2+(Y_\mathrm{cr}-y_\mathrm{k})^2}
\label{eqA3}
\end{equation} 
where $d_\mathrm{cr}$ is the distance of the RF detector from the EAS core. 


\end{document}